\documentclass{aa}

\usepackage{graphicx}
\usepackage{txfonts}
\usepackage{color}
\usepackage{natbib}

\begin{document}
    \title{Spheroidal core-mantle particle absorption, scattering, and polarisation in the long-wavelength limit}
    \subtitle{}

    \author{A.P. Jones\inst{1}
    \and
    N. Ysard\inst{2} 
    }
    
     \institute{Universit\'e Paris-Saclay, CNRS,  Institut d'Astrophysique Spatiale, 91405, Orsay, France.\\
               \email{anthony.jones@universite-paris-saclay.fr} 
            \and
                   IRAP, Universit\'e de Toulouse, CNRS, UPS, 9Av. du Colonel Roche, BP 44346, 31028 Toulouse, cedex4, France\\
              }

    \date{Received 2/10/2025; accepted 25/12/2025}


   \abstract
{The numerical calculation of optical properties (extinction, absorption, scattering and polarisation efficiencies) is often  time-consuming for non-spherical and inhomogeneous particles. Where possible analytical methods are therefore to be preferred.}
{We provide an analytical tool to derive the optical properties of mantled spheroidal particles, of arbitrary axis ratio, in the long wavelength limit ($a \ll \lambda$), where the mantle form may be confocal, co-axial or of constant depth with respect to the particle core. }
{We have developed an analytical approach to spheroidal core/mantle particle optical property calculations.}
{The analytical method compares well with DDSCAT numerical calculations and, under limited circumstances, with those made using the Bruggemann effective medium theory (EMT).} 
{The analytical method presented here provides a useful tool to explore the optical and polarisation properties of core/mantle spheroidal particles  at long wavelengths ($\lambda \gtrsim 8\mu$m) and is simpler and faster to implement than corresponding numerical methods.  We caution against the use of EMT methods in approximating the optical properties of core/mantle particles.}
   \keywords{ISM:abundances -- ISM:dust,extinction}

    \maketitle
    \nolinenumbers 
%

\section{Introduction}

Polarisation by interstellar and circumstellar grains is currently a topic of great interest in astronomy as the arrival of new data outstrips the ability of the current models to coherently explain these data. It is clear that the use of homogeneous, single-component particles is too simplistic \cite[{\it e.g.},][]{2011A&A...528A..98J} and it is perhaps this that most restricts our current understanding. Cosmic dust is indeed far from simple and  modelling realistic dust shapes (quasi-spherical, -spheroidal, -ellipsoidal, irregular, \ldots) and complex structures \citep[homogeneous, mantled, aggregate, aggregated core/mantled, etc., e.g.][]{2018A&A...617A.124Y,2019A&A...631A..88Y} can be computationally expensive, especially with tools such as the Discrete Dipole Approximation \citep[see for instance][DDSCAT]{2000ascl.soft08001D} or T-matrix \citep{1996JQSRT..55..535M}, depending on the required volume element resolution. A full investigation therefore involves a time-consuming and computationally heavy exploration of a very diverse parameter space \citep[wavelength, size, shape, structure, e.g.][]{2018A&A...617A.124Y,2019A&A...631A..88Y,2025A&A...698A.200C}. 

A theory, and the exact solutions, for the light scattering properties of homogeneous, isotropic, spheroidal particles was developed and evaluated about half a century ago by \cite{1975ApOpt..14...29A} and \cite{1979ApOpt..18..712A}, and that for the scattering and absorption properties of coated spheroids presented not long after by \cite{1980AnTok..18....1O} and \cite{vandeHulst}. 
With these methods it became possible to determine and explore the optical properties of homogeneous spheroids with size parameters $ \pi ( l / \lambda) < 30$  \cite[e.g.][]{1979ApOpt..18..712A}, where $l$ is a characterising particle dimension and $\lambda$ is the wavelength of the  incident radiation. Using the analytical solutions of \cite{1975ApOpt..14...29A}, \cite{1979ApJ...228..450R} investigated the extinction and polarisation produced by spheroidal particles and were able to put some constraints upon the likely shapes and refractive indices of interstellar grains by assuming an homogeneous, spheroidal dust model and good grain alignment. 

Following the logic of the THEMIS model \citep[The Heterogeneous dust Evolution Model for Interstellar Solids,][]{2013A&A...558A..62J,2015A&A...579A..15K,2017A&A...602A..46J,2024A&A...684A..34Y}, we do not enter into the full complexity of dust but, as a small step in this direction, consider the optical and polarising properties ($Q_i$, i = ext, sca, and abs, including their $||$ and $\perp$ components) of (hydrogenated) amorphous carbon \citep[a-C(:H),][]{2012A&A...540A...1J,2012A&A...540A...2J,2012A&A...542A..98J}\footnote{Aliphatic-rich and aromatic-rich hydrogenated amorphous carbon are designated a-C:H and a-C, respectively.} mantled spheroidal amorphous silicate  (a-Sil) particles, and core-mantle carbonaceous particles, using the derived optical constants of these materials \citep{2012A&A...540A...1J,2012A&A...540A...2J,2012A&A...542A..98J,2017A&A...606A..50D,2017A&A...600A.123D,2022A&A...666A.192D}. 

In this paper we begin by summarising and explicitly stating the necessary equations needed to derive the optical properties of core-mantle spheroidal particles. In this we were aided by the seminal works of \cite{vandeHulst} and \cite{1998asls.book.....B} and greatly in the debt of these authors. In the following we developed and extended the work of these authors to some more particular mantle shapes. 

This work is organised as follows: 
Section \ref{sect_eqns} outlines the fundamental equations for co-axial, confocal and constant depth mantled spheroids, 
Section \ref{sect_xsects} gives the extinction, absorption, scattering, and polarisation cross-sections for mantled spheroids, 
Section \ref{sect_results} presents the results of the calculations and comparisons with numerical calculations made with DDSCAT, effective medium theory volume averaging methods, and 
Section \ref{sect_conclusions} summarises this work and concludes.

\section{The equations for mantled spheroids}
\label{sect_eqns}

\begin{figure}
\centering
\includegraphics[width=9.0cm]{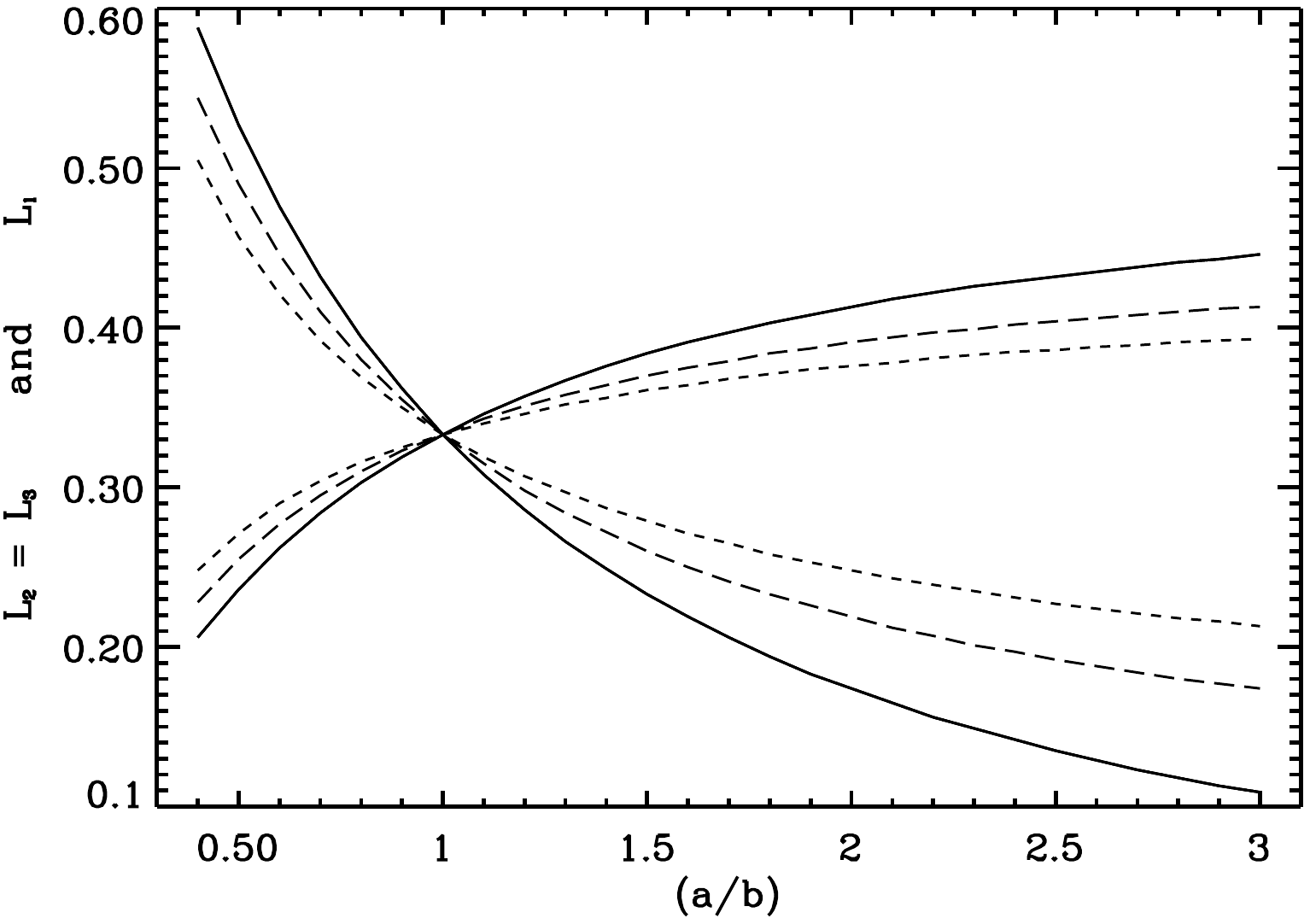}
\caption{Co-axial, confocal, and constant depth mantled spheroidal particle $L_i$ factors for an $a$ axis to mantle depth ratio $a/d \simeq 3$. The solid lines show the behaviour for the cores and core+mantle for the co-axial case. The short (long) dashed line show the  cores+mantle $L_i$ factors for the confocal (constant depth) case. The $L_1$ factors decrease with increasing $a/b$, while those for $L_2$ and $L_3$ increase.}
\label{fig_Ls}
\end{figure}

\begin{figure}
\centering
\includegraphics[width=8.5cm]{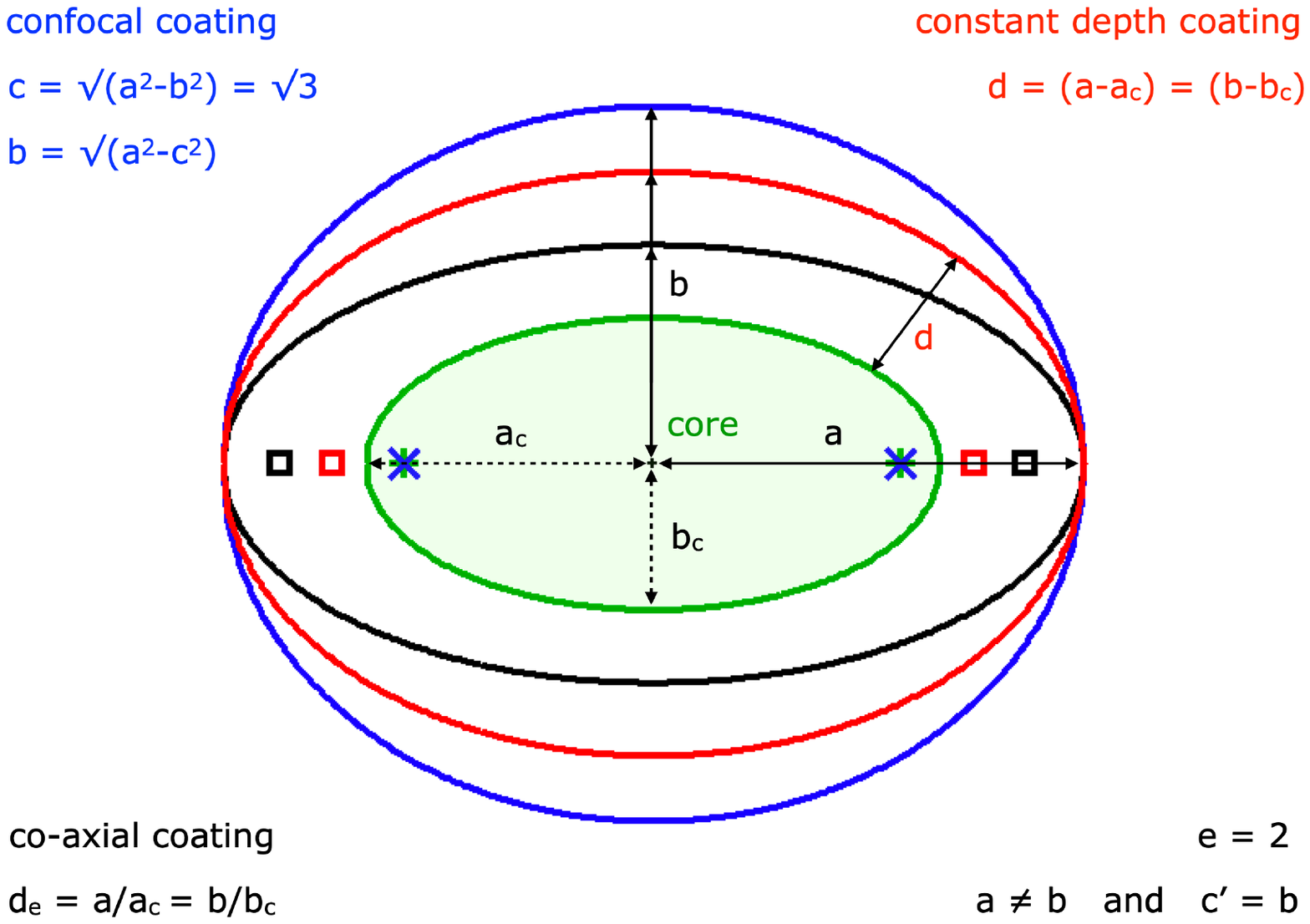}
\caption{Confocal (blue), constant depth (red) and co-axial (black) coated or mantled spheroidal cores (green) shown in cross-section. The spheroid semi-axes are $a$, $b$ and $c^\prime$, where the $c^\prime$ axis is perpendicular to the plane of the figure. For the confocal case the linear eccentricity $c$ $= \surd (a^2-b^2)$. The coloured symbols indicate the foci of the elliptical cross-sections for the core and core+mantle.}
\label{fig_cellips}
\end{figure}

We first consider the general equations for un-mantled or bare ellipsoids and spheroids with sole focus on the latter. 
In ellipsoidal  particles the three semi-major axes ($a$, $b$ and $c$) are unequal ($a \neq b \neq c$) and the particle shape and volume, $V$, are defined, with reference to the orthogonal Cartesian axes $x$, $y$, and $z$, by 
\begin{equation}
\frac{x^2}{a^2} + \frac{y^2}{b^2} +\frac{z^2}{c^2} = 1 \ \ \ \ \ {\rm and} \ \ \ \ \ 
V = \frac{4}{3} \pi \left( \, a \, b \, c \, \right), 
\end{equation} 
respectively. The determination of the surface area is considerably more complex (see later). In the simpler case of spheroidal particles with two of the semi-major axes of equal length, where $a \neq b$ and  $c = b$, the shape and volume are defined by
\[
\frac{x^2}{a^2} + \frac{1}{b^2}(y^2+z^2) = 1   \ \ \ \ \ {\rm or} \ \ \ \ \ x^2 + \left( \frac{a}{b} \right)^2(y^2+z^2) = a^2 \ \ \ \ \ {\rm and} \ \ \ \ \ 
\]
\begin{equation}
V = \frac{4}{3} \pi \left( \, a \, b^2 \, \right), 
\end{equation} 
respectively.  The eccentricities for prolate ($e_p$, $a>b$) and oblate spheroids ($e_o$, $a<b$) are,  
\begin{equation}
e_p = \left( 1 - \frac{b^2}{a^2} \right)^{\frac{1}{2}} \ \ \ \ \ {\rm and} \ \ \ \ \ e_o = \left( \frac{b^2}{a^2} - 1 \right)^{\frac{1}{2}}.
\end{equation}
Given that we no longer need to consider the $c$ axis in our equations for spheroids, by convention we use $c$ for the linear eccentricity $c = \surd(a^2-b^2)$.  The third semi-major axis is now labelled $c^\prime$ when referred to in any of the following equations. 

For spheroidal particles the shape-dependent factors $L_i$ ($i = 1, 2$ or 3) used in the calculation of the particle polarisability sum to unity, that is, 
\begin{equation}
\sum_{i = 1,2,3} L_i = 1.  
\end{equation}
For spherical particles ($a = b = c^\prime$) the shape factors along the three orthogonal particle axes are equal, that is $L_1 = L_2 = L_3 = \frac{1}{3} $. For prolate spheroids  ($a > b$ and $b = c^\prime$) the shape factor, $L_1$, factor along the $a$ axis, the axis of symmetry, is given by 
\begin{equation}
L_{1} = \frac{ ( 1 - e_p^2 ) }{ e_p^2 } \left[  -1 + \frac{1}{2 e_p} {\rm log} \left\{ \frac{(1+e_p)}{(1-e_p)} \right\} \right], 
\end{equation} 
and for oblate spheroids ( $a < b$ and $b = c^\prime$) the shape factor, $L_1$, again along the axis of symmetry, is equivalently 
\begin{equation}
L_1 = \frac{ (1+e_o^2)}{e_o^2} \left[ 1 - \frac{1}{e_o} \, {\rm arc\ tan}( \, e_o \, ) \right]. 
\end{equation}
Here the \cite{vandeHulst} semi-axis conventions have been followed rather than those of \cite{1998asls.book.....B}.
For both prolate and oblate spheroids the $L_2$ and $L_3$ factors are simply $L_2 = L_3 = \frac{1}{2} ( 1 - L_1 )$. For coated or mantled spheroidal particles the shape factors,  $L_i$, have the same mathematical form for the particular type of spheroid core and mantle under consideration\footnote{N.B. This is strictly only valid where both the core and mantle are spheroidal. For constant depth mantles on spheroidal cores this is no longer true because the mantle is strictly not spheroidal, even though its surface is extremely close to spheroidal in the thin-mantle case.} but are obviously numerically different where the core and mantle do not have the same eccentricity. From here on the particle coating will be referred to as a mantle in order to avoid subscript confusion with the particle core and these inhomogeneous structures are from here on labelled core-mantle or CM particles. Following the THEMIS convention for CM particles,  amorphous carbon mantles on amorphous silicate cores and on hydrogenated amorphous carbon cores are designated a-Sil/a-C and a-C:H/a-C, respectively. 

The $L_i$ factors as a function of $(a/b)$ are shown in Fig. \ref{fig_Ls} for the co-axial, confocal, and constant depth mantled spheroidal particle cases for a mantle depth ratio $(a/d) \simeq 3$, that is for a relatively thick mantle in order to accentuate the difference between the core and total (core+mantle) factors. In the co-axial case, the ratio of the outer radius to that of the core is fixed in all directions; the $L_i$ factors for core and mantle are therefore the same. As shown in Fig. \ref{fig_cellips} confocal mantles tend to make the particles more spherical, as do constant depth mantles but less so than for the confocal form. Note that with increasing $a/b$ the $L_1$ factors decrease while $L_2$ and $L_3$ increase. Of particular note is that, in the constant depth and confocal mantle cases, the core and mantle $L_i$ factors are offset, an effect that is more pronounced with the confocal form.  

We now concentrate on the mantled forms of spheroidal particles, hereafter referred to as cores, with semi-axes $a_c$ and $b_c$, and mantles of (non)uniform depth, $d$, which do not necessarily have the same eccentricities as the core. In Fig.~\ref{fig_cellips} we show the  three different mantle depth cases that are considered here:
\begin{itemize}
\item co-axial: with constant mantle to core axis ratio, $d_e$,   
\item confocal: where the core and mantle have the same linear eccentricity $c$, and 
\item constant depth: where the mantle depth, $d$, normal to the surface is constant.
\end{itemize}
The details of each of the three mantle forms are presented in the following subsections.

\subsection{Co-axial mantled spheroids}
\label{sect_d_co-ax}

For co-axial CM spheroidal particles the ratio of core+mantle semi-axis to the core semi-axis, $d_e$, is fixed, that is    
\begin{equation}
d_e = \left( \frac{a}{a_c} \right) = \left( \frac{b}{b_c} \right)
\end{equation} 
and the mantle shape is defined by the equation: 
\[
\frac{x^2}{( a_c \, d_e )^2} + \frac{1}{(b_c \, d_e)^2}(y^2+z^2) = 1   \ \ \ \ \ {\rm or} \ \ \ \ \ 
\]
\begin{equation}
x^2 + \frac{( a_c \, d_e )^2}{(b_c \, d_e)^2} (y^2+z^2) = (a_c \, d_e )^2 = a^2. 
\end{equation}
The core, total (core + mantle), and mantle volumes, $V_c$, $V_t$, and $V_m$, respectively, are given by  
\[
V_c = \frac{4}{3} \pi \ ( \, a_c \, b_c^2 \, ) = \frac{4}{3} \pi \  \frac{( \, a \, b^2 \, )}{d_e^3},  \ \ \ \ V_t = \frac{4}{3} \pi \left( \, a \, b^2 \, \right),   \ \ \ \ \ {\rm and} \ \ \ \ \ 
\]
\begin{equation}
V_m = \left( V_t - V_c \right) = \frac{4}{3} \pi \left( \, a \, b^2 \, \right) \left( \, 1 - \frac{1}{d_e^{3}} \, \right). 
\end{equation} 
In co-axial spheroids, where the ratio $d_e$ is same in all directions, the mantle depth depends on the radial direction, being thickest along the major axis $a$ and thinnest along the minor $b$ and $c^\prime$ axes for prolate spheroids (see Fig.~\ref{fig_cellips}). The opposite is true for oblate spheroids. For co-axially mantled prolate spheroids ($a > b$) where the core and the mantle have the same axis ratio, that is $d_e = b_c/a_c = b/a$, the eccentricity, $e_p$, in the $L$ factors ($L_1$, $L_2 = L_3$) are the same for the particle core and its mantle, 
\begin{equation}
e_p^2 = \left[ 1 - \left( \frac{b}{a} \right)^2 \right] = \left[ 1 - \left( \frac{b_c}{a_c} \right)^2 \right]. 
\end{equation} 
For co-axially-coated oblate spheroids ($a<b$) the eccentricity, $e_o$, in the $L$ factors ($L_1$, $L_2 = L_3$) are, as for co-axially-coated prolate spheroids, identical for the particle core and mantle,   
\begin{equation}
e_o^2 = \left[ \left( \frac{b}{a} \right)^2 - 1 \right]  = \left[ \left( \frac{b_c}{a_c} \right)^2 - 1 \right]. 
\end{equation}

\subsection{Confocal mantled spheroids}
\label{sect_d_confocal}

For confocal spheroidal particles, where the foci of the core and mantle coincide, that is they have the same linear eccentricities, $c$, the minor semi-axis lengths can be expressed as
\begin{equation}
b_c = \surd (a_c^2-c^2)  \ \ \ \ \ {\rm and} \ \ \ \ \  b = \surd (a^2-c^2).  
\label{eq_c}
\end{equation}
The mantle depths along the major and minor axes, $( a - a_c )$ and $( b - b_c )$, respectively, are unequal as are the ratios of the core and mantle semi-axis ratios, that is $(a/a_c) \neq (b/b_c)$. In this case the core shape is defined by the equation: 
\[
\frac{x^2}{a_c^2} + \frac{y^2}{(a_c-c)^2} + \frac{z^2}{(a_c-c)^2} = 1   \ \ \ \ \  {\rm or} \ \ \ \ \ 
\]
\begin{equation}
x^2 + \frac{a_c^2}{(a_c-c)^2} (y^2+z^2) = a_c^2 
\label{eq_definition_1}
\end{equation} 
and the mantle shape by 
\[
\frac{x^2}{a^2} + \frac{y^2}{(a-c)^2} + \frac{z^2}{(a-c)^2} = 1   \ \ \ \ \ {\rm or} \ \ \ \ \ 
\]
\begin{equation}
x^2 + \frac{a^2}{(a-c)^2} (y^2+z^2) = a^2.  
\label{eq_definition_2}
\end{equation} 
The core, total, and mantle volumes, $V_c$, $V_t$, and $V_m$, respectively, are given by 
\[
V_c = \frac{4}{3} \pi \  ( \, a_c \, b_c^2 \, ) = \frac{4}{3} \pi \  \left[ \, a_c \, (a_c^2-c^2) \, \right], \ \ \ \ \  
\]
\[
 V_t = \frac{4}{3} \pi \left[ \, a \, (a^2-c^2) \, \right], \ \ \ \ \ {\rm and} \ \ \ \ \ 
 \]
 \begin{equation}
 V_m = \left( V_t - V_c \right) = \frac{4}{3} \pi \left[ \, a^3 - c^2 \left( a - a_c \right) -  a_c^3 \, \right].  
\label{eq_Vs}
\end{equation}
The linear eccentricity, $c$, is real for $a_c \geqslant b_c$ and $a \geqslant b$, and consequently Eq.~(\ref{eq_c}) is therefore strictly only valid for prolate spheroids. However, given that $c^2$ is negative for oblate spheroids and always a subtractive term in Eqs.~(\ref{eq_definition_1}) to (\ref{eq_Vs}), Eq.~(\ref{eq_Vs}) does yield valid volumes for all mantled spheroidal grains.\footnote{The only inconvenience is that we can no longer plot the foci for oblate spheroids because they fall in the complex domain.} In such spheroids the mantle depth depends on the radial direction, being thicker along the minor axes in the case of prolate spheroids (see Fig.~\ref{fig_cellips}). For oblate spheroids the mantle is thicker along the major axis. For confocally-mantled prolate spheroids the mantle and core eccentricities, $e_{p,m}$ and $e_{p,c}$, are 
\[
e_{p,m}^2 = \left[ 1 - \left( \frac{a^2-c^2}{a^2} \right) \right]  = \left( \frac{c}{a} \right)^2 \ \ \ \ \ {\rm and} \ \ \ \ \  
\]
\begin{equation}
e_{p,c}^2 = \left[ 1 - \left( \frac{a_c^2-c^2}{a_c^2} \right) \right]  = \left( \frac{c}{a_c} \right)^2. 
\end{equation}
In the case of confocally-mantled oblate spheroids the eccentricities for the entire particle and the core, $e_{o,m}$ and $e_{o,c}$, are 
\[
e_{o,m}^2 = \left[ \left( \frac{a^2-c^2}{a^2} \right) - 1 \right]  = -\left( \frac{c}{a} \right)^2 \ \ \ \ \ {\rm and} \ \ \ \ \  
\]
\begin{equation}
e_{o,c}^2 = \left[ \left( \frac{a_c^2-c^2}{a_c^2} \right) - 1 \right]  = -\left( \frac{c}{a_c} \right)^2. 
\end{equation}
Again the negative values do not pose a problem because $c^2$ is negative for oblate spheroids.

\subsection{Constant depth mantled spheroids}
\label{sect_d_const}

In the interstellar medium the physical processes of accretion (of gas phase species or of nanoparticles), photo-processing in an isotropic radiation field or erosion by sputtering all result in the formation or erosion of constant depth surface layers. Thus, this would appear to be the case most appropriate for the astrophysical situation. It is, however, also the most complex of the three cases. This is because in constant depth coated core/mantle particles either the core or the mantle is spheroidal while the other, although it may be extremely close to exhibiting a spheroidal surface in the thin-mantle limit, is not exactly spheroidal. Thus, for coated spheroidal particles where the mantle depth, $d$, is constant there exist two possible cases and for both 
\begin{equation}
d = ( a - a_c ) = ( b - b_c ) 
\end{equation}
because this expression includes only the core semi-axes and the mantle depth along those semi-axes, it says nothing of the intermediate behaviours. The two possible cases are subtly rather different because either $a$ and $b$ or $a_c$ and $b_c$ define a spheroid. Given the need to directly compare the three CM cases, and that the mantle is generally thin compared to the core size, as in the large grains in the THEMIS model (viz., $5-20$nm-thick mantles on $130 - 150$nm radius cores), only the spheroidal core case is considered here. The core is then taken to have an exact spheroidal form with its shape and volume, $V_c$, determined by the following equations:
\[
\frac{x^2}{a_c^2} + \frac{1}{b_c^2} (y^2+z^2)  = 1 \ \ \ \ \ {\rm or} \ \ \ \ \ 
\]
\begin{equation}
x^2 + \frac{a_c^2}{b_c^2}(y^2+z^2) = a_c^2, \ \ \ \ \ 
V_c = \frac{4}{3} \pi \  ( \, a_c \, b_c^2 \, ). 
\end{equation} 
This is where, as referred to above, things get interesting. The mantle volume can be exactly determined using Steiner's formula\footnote{In the confectionery industry it turns out that this formula is critical in determining the quantity of coating required to cover an M\&M. }  for the volume of a coating on any convex shape where the mantle depth, normal to the core surface, is the same at each point on that surface, and is given by 
\begin{equation}
V_{\rm coat} = \ S_c \ d \ + \ \pi \ l \ d^2 \ + \ \frac{4}{3} \pi \ d^3
\label{eq_Steiner}
\end{equation}
where $S_c$ is the spheroidal core surface area, $l$ is the mean length and $d$ is the mantle depth. The first term provides the thin mantle limit ($d \ll a_c$, i.e. volume = area $\times$ depth) and the third term is the thick mantle limit ($d \gg a_c$, i.e. for a spherical particle). The second term is returned to below after we consider the surface area calculation for a spheroid. The surface areas of prolate and oblate spheroidal cores, $S_{c,p}$ and $S_{c,o}$, respectively, are given by
\[
S_{c,p} = 2 \pi \, b_c^2 \left( 1 + \frac{a_c}{b_c \ e_p} \ {\rm arcsin} \ e_p \right)
 \ \ \ \ \ {\rm and} \ \ \ \ \ 
\]
\begin{equation}
S_{c,o} = 2 \pi \, b_c^2 + \pi \, \frac{a_c^2}{\epsilon_o} \ {\rm log} \left( \frac{1 + \epsilon_o }{1 - \epsilon_o}  \right) \ \ {\rm where} \ \ \epsilon_o^2 = 1 - \frac{a_c^2}{b_c^2}. 
\end{equation}
For consistency with the above-defined eccentricity, $e_o$, where $\epsilon_o = (a_c/b_c) \, e_o$, the equation for $S_{c,o}$ can be re-written as 
\begin{equation}
S_{c,o} = 2 \pi \, b_c^2 + \pi \ \frac{ a_c \, b_c }{ e_o } \ {\rm log} \left[ \frac{ (b_c/a_c) + e_o }{ (b_c/a_c) - e_o }  \right].  
\end{equation}
Returning to the second term in Eq.~(\ref{eq_Steiner}), this is indeed more complex but, geometrically, the mean length, $l$, in Eq.~(\ref{eq_Steiner}) is a mean curvature (an inverse length) times a surface area. For a spheroid ($a_c \geqslant b_c \geqslant c_c^\prime$) this can be calculated using the following equation (see links in the following related footnotes):
\[
l(a_c,b_c,c_c^\prime) = \frac{1}{2\pi} \int_0^{2\pi} \int_0^\pi {\rm sin} \, \theta \ \surd ( [ a_c^2 \, {\rm cos}^2  \phi + b_c^2 \, {\rm sin}^2  \phi] \, {\rm sin}^2 \, \theta 
\ldots
\]
\begin{equation}
\ldots + (c_c^\prime)^2 \, {\rm cos}^2 \, \theta \, ) \ d\theta \ d\phi, 
\label{eq_mean_length}
\end{equation}
which can be solved numerically. Hence, in the results presented here the mantle volume was calculated using a lightly-modified version of the {\tt coating.f90} code made public by F. X. Timmes.\footnote{http://cococubed.asu.edu/code\_pages/coating.shtml \\ The only modifications made to {\tt coating.f90} were to the input and output functions for compatibility with the routines developed here.}
Alternatively, using Thomsen's formula, the ellipsoid surface area, $S_{\rm ell}$, can be approximated to within $\approx 1$\% by
\begin{equation}
S_{\rm ell} = 2 \pi \ \left( a_c^p b_c^p + a_c^p c_c^p + b_c^p (c_c^\prime)^p \right)^{1/p},  
\label{eq_Sabc}
\end{equation}
where setting $p=1.6075$ leads to a relative error of less than $1.42$\% (Thomsen, 2004: see footnote) or with $p = 8/5$ to a maximum error of $-1.178$\% (Cantrell, 2004: see footnote).\footnote{http://numericana.com/answer/ellipsoid.htm\#ellipsoid} The particle core, mantle and total (core + mantle) volumes, $V_c$, $V_m$ and $V_t$, respectively, are then
\begin{equation}
V_c = \frac{4}{3} \pi \  ( \, a_c \, b_c^2 \, ),  \ \ \ \ \  V_m =  V_{\rm coat}, \ \ \ \ \ {\rm and} \ \ \ \ \ V_t = V_c + V_{\rm coat}.
\end{equation} 
For constant mantle depth, $d$, coated prolate spheroids, the mantle and core eccentricities, $e_{p,m}$ and $e_{p,c}$, are 
\begin{equation}
e_{p,m}^2 = \left[ 1 - \left( \frac{b_c+d}{a_c+d} \right)^2 \right]  \ \ \ \ \ {\rm and} \ \ \ \ \  e_{p,c}^2 = \left[ 1 - \left( \frac{b_c}{a_c} \right)^2 \right]. 
\end{equation}
In the case of constant depth-coated oblate spheroids the eccentricities for the mantle and core, $e_{o,m}$ and $e_{o,c}$, are 
\begin{equation}
e_{o,m}^2 = \left\{ \left[ \frac{(b_c+d)}{(a_c+d)} \right]^2 - 1 \right\}  \ \ \ \ \ {\rm and} \ \ \ \ \  e_{o,c}^2 = \left[ \left( \frac{b_c}{a_c} \right)^2 -1 \right]. 
\end{equation}

\subsection{Mantle volume normalisation}

A comparison of the different mantle cases was made using equal volume mantles, as defined by the mantle volume in the constant depth mantle case, the most mathematically demanding of the three scenarios. Hence, the mantle depth, $d$, is first defined and the mantle material volume, $V_m$, calculated as described above (Section~\ref{sect_d_const}). This volume is then used to determine the co-axial case mantle  semi-axis ratio, $d_e$, (Section~\ref{sect_d_co-ax}), 
\begin{equation}
d_e = \left( \ \frac{3}{4 \pi} \ \frac{V_m}{a_c \, b_c^2 } + 1 \right)^{\frac{1}{3}} = \left( \frac{V_t}{V_m} \right)^{\frac{1}{3}}, 
\end{equation}
which is then in turn used to determine the core and mantle linear eccentricities, $c$, in the confocal core-mantle case (Section~\ref{sect_d_confocal}) as a function of the as yet to be determined confocal case mantle semi-axis ratio, $d_{ec}$, where 
\[
c = \surd( a_c^2 - b_c^2 ),  \ \ \ \ \  a = a_c \, d_{ec},  \ \ \ \ \ {\rm and} \ \ \ \ \ 
\]
\begin{equation}
b = \surd( a^2 - c^2 ) = \surd( a_c^2 \, d_{ec}^2 - c^2 ). 
\end{equation}
Given that the mantle volume in the confocal case can be expressed as 
\begin{equation}
V_m = \frac{4}{3} \pi \left( \, a_c^3 \, d_{ec}^3 - a_c \, c^2 \, d_{ec} + a_c \, c^2 -  a_c^3 \, \right)  
\label{eq_Vconf}
\end{equation}
finding $d_{ec}$ requires solving a cubic equation of the form $d_{ec}^3 + p \, d_{ec} + q = 0$, that is, 
\[
0 = d_{ec}^3 - \left( \frac{c}{a_c}\right)^2  d_{ec} + \left[ \, \left( \frac{c}{a_c}\right)^2 - \frac{3 \, V_m}{4 \pi \, a_c^3} - 1 \right],  
 \ \ \ \ \  
\]
 \begin{equation}
{\rm with} \ \ \ \ \ p = - \left( \frac{c}{a_c}\right)^2  \ \ \ \ \ {\rm and} \ \ \ \ \ q = \left[ \, \left( \frac{c}{a_c}\right)^2 - \frac{3 \, V_m}{4 \pi \, a_c^3} - 1 \right].
\end{equation}
Using Cardano's formula this cubic equation has a real solution
\begin{equation}
d_{ec} =  C - \frac{p}{3C} \ \ \ \ \ {\rm where} \ \ \ \ \ 
C = \left[ -\frac{q}{2} + \left( \frac{q^2}{4} + \frac{p^3}{27}\right)^{\frac{1}{2}} \right]^{\frac{1}{3}}. 
\end{equation}

\section{The extinction, absorption and scattering cross-sections of mantled spheroids}
\label{sect_xsects}

\begin{figure*}
\centering
\includegraphics[width=18.0cm]{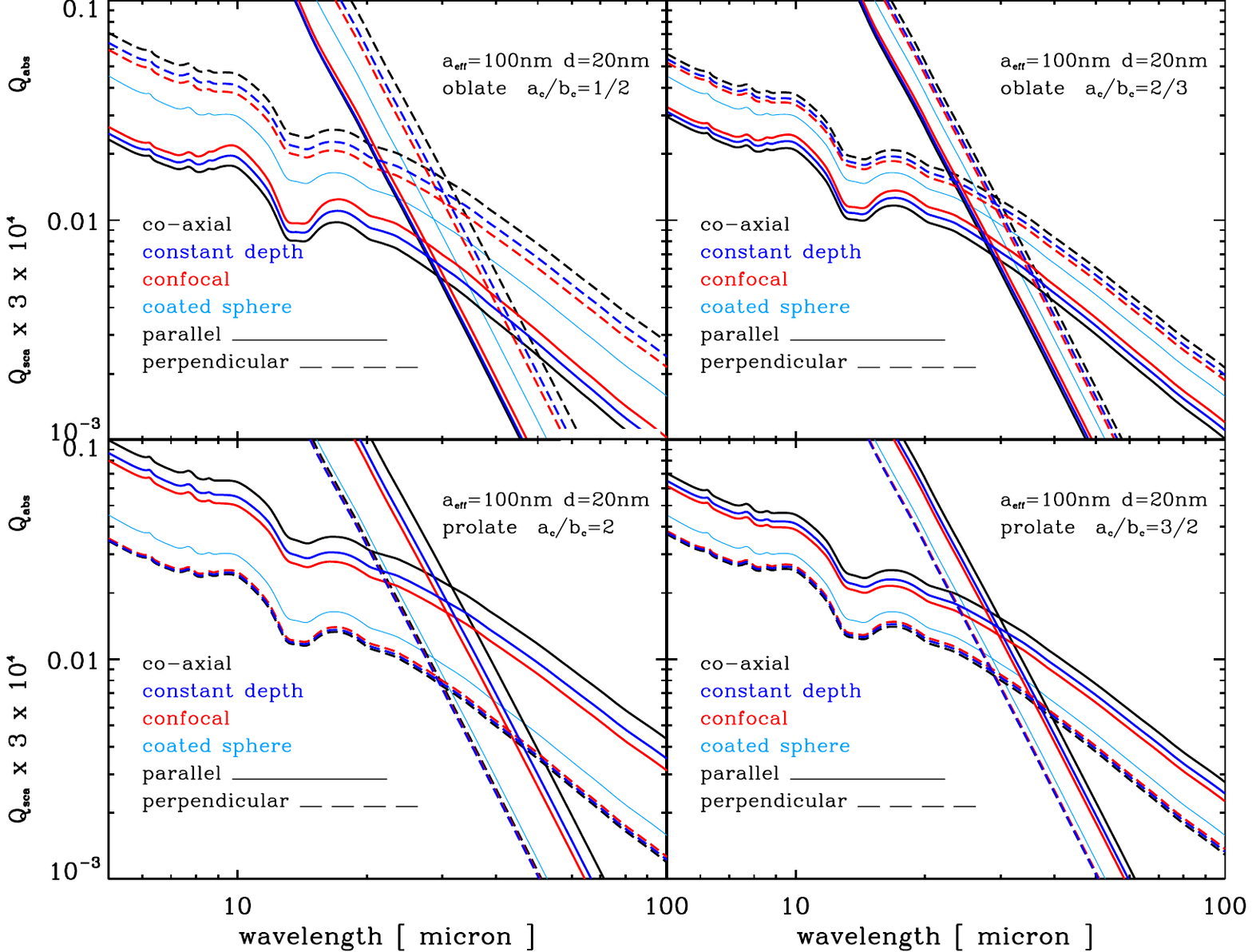}
\caption{Spheroid and coated sphere $Q_{\rm abs}$ and $Q_{\rm sca}$ factors, the latter multiplied by $3 \times 10^4$ for clarity, for the parallel (solid lines) and perpendicular (dashed lines) cases, normalised to the same core and mantle volumes. The data shown are for amorphous silicate (a-Sil) cores (effective radius 80nm) with 20nm deep aromatic-rich carbon (a-C) mantles (assumed radius 10nm, see Section \ref{sect_results}), as a function of wavelength and $e=(a_c/b_c)=0.5, 0.667, 1, 1.5,$ and 2 for co-axial (black), constant depth (blue) and confocal (red) mantles. Also shown for comparison are the data for mantled spheres (cobalt) of outer effective radius $a_{\rm eff} = 80 + 20 = 100$nm.} 
\label{fig_test_two}
\centering
\end{figure*}

For a coated spheroidal particle, with core material dielectric function $\varepsilon_c$ and mantle material dielectric function $\varepsilon_m$,  embedded in a medium with dielectric function $\varepsilon_v$  (for vacuum $\varepsilon_v = 1$) the polarisability of the particle as given by \cite{1998asls.book.....B} is 
\[
\alpha_i = \frac{ (V_t  / 4 \pi ) \ \{ ( \varepsilon_m - \varepsilon_v ) \ [ \varepsilon_m + ( \varepsilon_c - \varepsilon_m ) \ ( L_{c,i} - (V_c/V_t) \ L_{m,i} ) ]  }{ [ \varepsilon_m + ( \varepsilon_c - \varepsilon_m ) \ ( L_{c,i} - (V_c/V_t) \ L_{m,i} ) ][ \varepsilon_v + ( \varepsilon_m - \varepsilon_v ) \ L_{m,i}  ]} \ldots
\]
\begin{equation}
\ \ \ \ \ \ \ \ \ \ldots \frac{+ (V_c/V_t) \ \varepsilon_m \ ( \varepsilon_c - \varepsilon_m ) \} } 
 {+ (V_c/V_t) \ L_{m,i} \ ( \varepsilon_c - \varepsilon_m )} 
\end{equation} 
where $i = 1$, 2 or 3, and $\alpha_2 = \alpha_3$. When the core and mantle are of the same material, $\varepsilon_c = \varepsilon_m = \varepsilon$, this reduces to 
\begin{equation}
\alpha_i = \frac{ (V_t / 4 \pi )  \ ( \varepsilon - \varepsilon_v) \ \varepsilon } { \varepsilon \ [ \varepsilon_v + ( \varepsilon - \varepsilon_v ) L_{i} ] } = (V_t / 4 \pi )  \ \left[ L_{i} + \varepsilon_v \, (\varepsilon - \varepsilon_v)^{-1} \right]^{-1},
\end{equation}
the equation for an homogeneous spheroid. For each of the semi-major axes of a spheroid, $i = 1$, 2 and 3, with 1 parallel to the axis of symmetry and 2 and 3 perpendicular to that axis and therefore equal, the absorption and scattering efficiency factors, with absorption generally dominating over scattering at long wavelengths, are given by 
\[
Q_{\rm abs,i} = - 4 x  \ {\rm Im}( \, \alpha_i \, )   \ \ \ {\rm and} \ \ \ 
Q_{\rm sca,i} = \frac{8}{3} x^4  \ {\rm abs} ( \, \alpha_i^2 \, ),  \ \ \ \ \ 
\]
\begin{equation}
{\rm where} \ \ \ \ \ x = \frac{2 \pi }{\lambda} \frac{( a + 2 b )}{3}.   
\end{equation} 
The absorption cross-sections, with respect to the particle symmetry axis, in the parallel ($\|$) and perpendicular ($\perp$) directions are then 
\[
\sigma_{\rm abs,\|} = \pi \, b^2 \, Q_{\rm abs,1}  \ \ \  {\rm and} \ \ \ 
\]
\begin{equation}
\sigma_{\rm abs,\perp} = \pi \, (ab) \, Q_{\rm abs,2} = \pi \, (ab) \, Q_{\rm abs,3}. 
\end{equation} 
Note that in each case the relevant cross-section is the cross-section projected along that axis. Similarly, the respective scattering cross-sections in the $\|$ and $\perp$ directions are 
\[
\sigma_{\rm sca,\|} = \pi \, b^2 \, Q_{\rm sca,1}  \ \ \ {\rm and} \ \ \ 
\]
\begin{equation}
\sigma_{\rm sca,\perp} = \pi \, (ab) \, Q_{\rm sca,2} = \pi \, (ab) \, Q_{\rm sca,3},  
\end{equation} 
and the extinction cross-sections in the $\|$ and $\perp$ directions are 
\[
\sigma_{\rm ext,\|} = \pi \, b^2 \, ( Q_{\rm abs,1}  + Q_{\rm sca,1} ) \ \ \ \ \ {\rm and} \ \ \ \ \ 
\]
\begin{equation}
\sigma_{\rm ext,\perp} = \pi \, (ab) \, ( Q_{\rm abs,2} + Q_{\rm sca,2}) = \pi \, (ab) \, ( Q_{\rm abs,3} + Q_{\rm sca,3}).  
\end{equation} 
The parallel and perpendicular absorption and scattering efficiency factors, for the three mantle shapes are compared in Fig. \ref{fig_test_two}. The equally-weighted, semi-major axis-averaged absorption, scattering and extinction cross-sections can be expressed as  
\[
\langle \sigma_{\rm abs} \rangle = \frac{1}{3} ( \sigma_{\rm abs,\|} + 2 \ \sigma_{\rm abs,\perp} ), \ \ \  
\langle \sigma_{\rm sca} \rangle = \frac{1}{3} ( \sigma_{\rm sca,\|} + 2 \ \sigma_{\rm sca,\perp} ),    \ \ \ 
\]
\begin{equation}
{\rm and} \ \ \ \langle \sigma_{\rm ext} \rangle = \langle \sigma_{\rm abs} \rangle + \langle \sigma_{\rm sca} \rangle.  
\label{eq_sigma_t}
\end{equation} 

\begin{figure*}
\centering
\includegraphics[width=18.0cm]{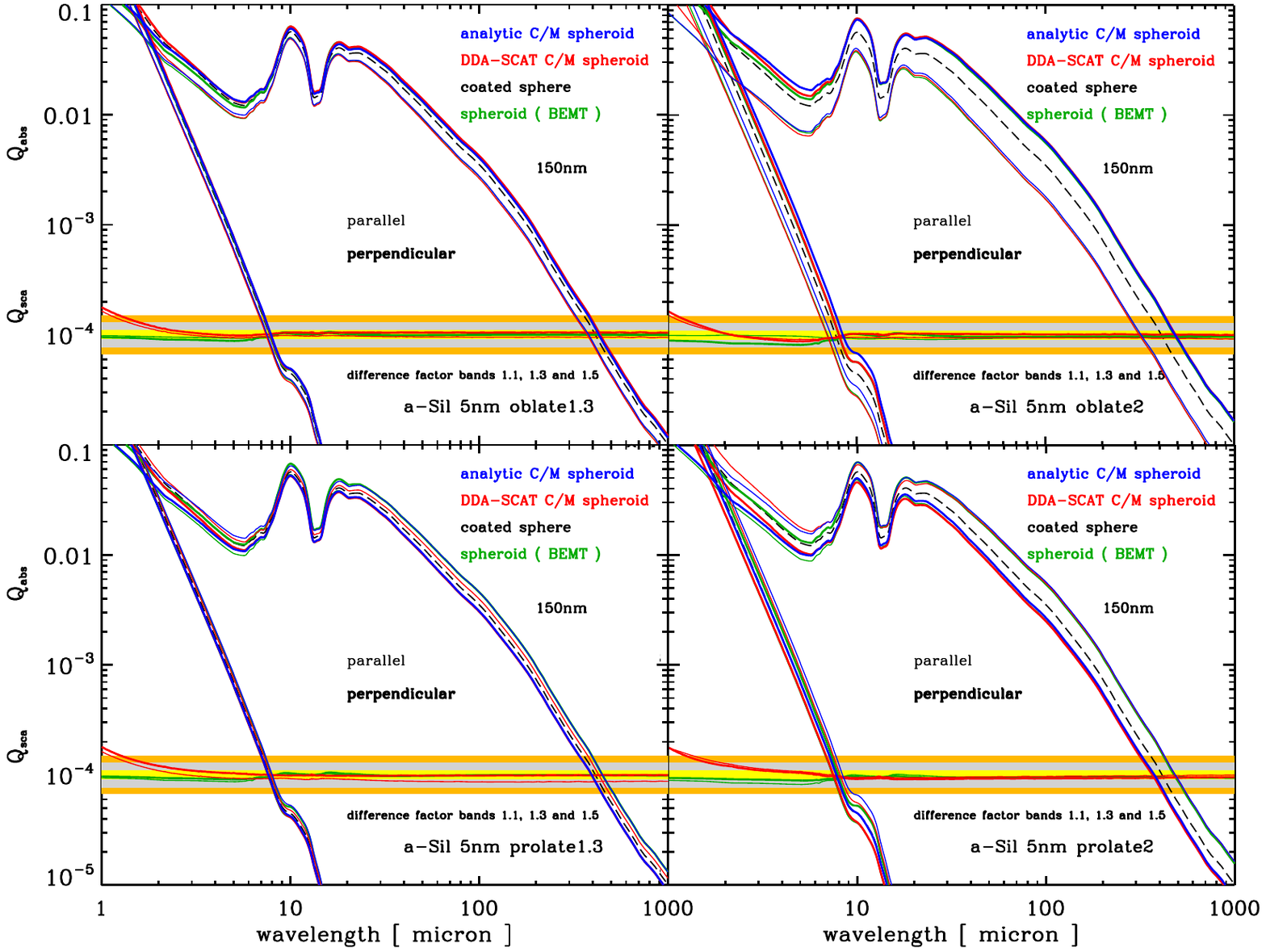}
\caption{Optical properties, $Q_{\rm abs}$ and $Q_{\rm sca}$, for spheroid and coated spheroid, parallel (thin lines) and perpendicular (thick lines), for amorphous silicate (a-Sil) particles of $a_{\rm eff} =150$nm with 5nm deep a-C confocal mantles; for oblate and prolate particle axis ratios of 1.3 and 2. The key parameters are indicate on each sub-plot. The horizontal bands in the lower part of each plot show the ratio of the $Q_{\rm abs}$ values for DDA/analytic (red) and EMT/analytic (green). The horizontal bands delimit difference factor ranges of $0.9-1.1$ (yellow), $0.7-1.3$ (grey) and $0.5-2$ (orange).} 
\label{fig_aSil_150}
\centering
\end{figure*}

\begin{figure*}
\centering
\includegraphics[width=18.0cm]{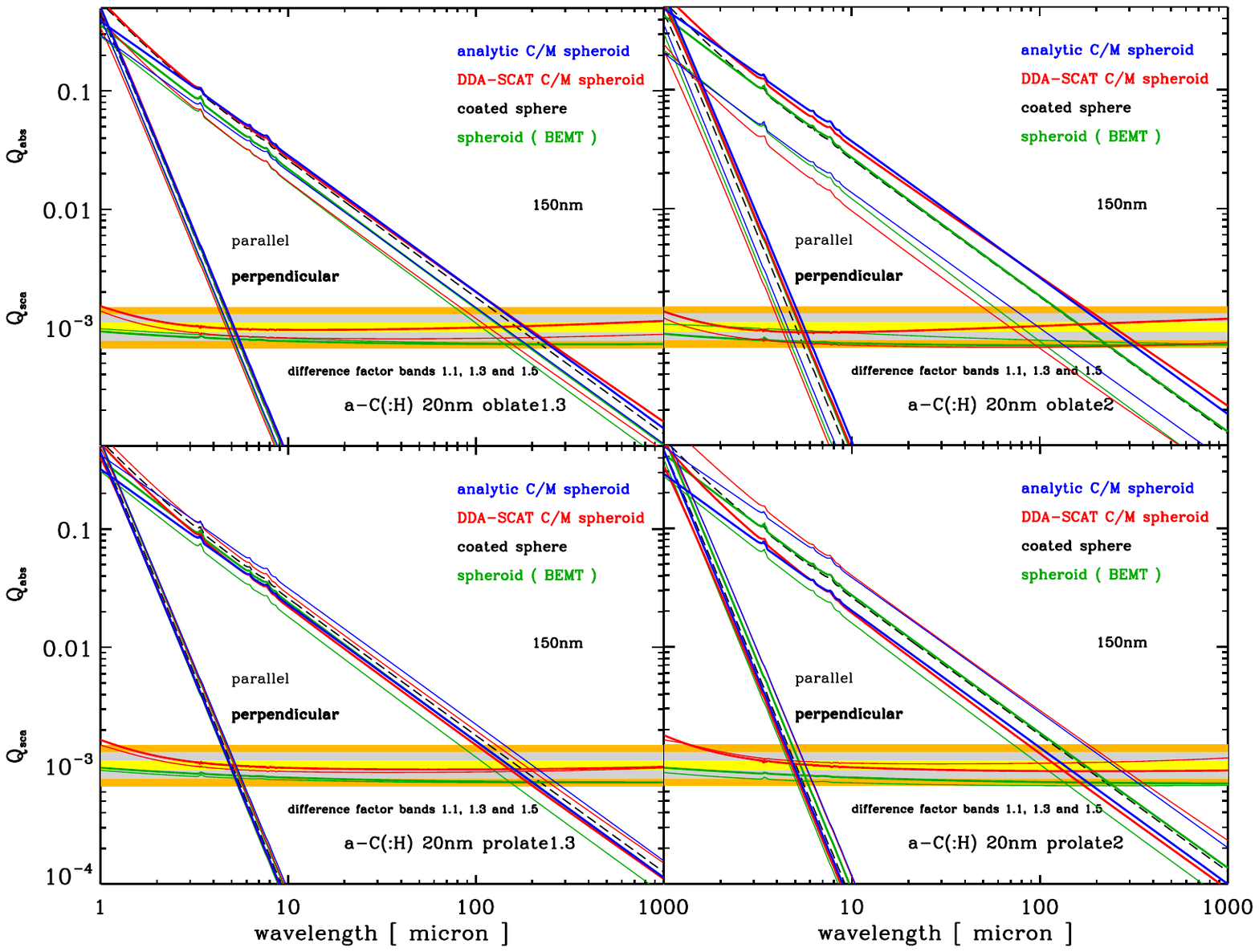}
\caption{Same as Fig. \ref{fig_aSil_150} but for amorphous carbon (a-C:H) particles with $a_{\rm eff} =150$nm with 20nm deep confocal a-C mantles.} 
\label{fig_aCH_150}
\centering
\end{figure*}

\begin{figure}
\centering
\includegraphics[width=8.5cm]{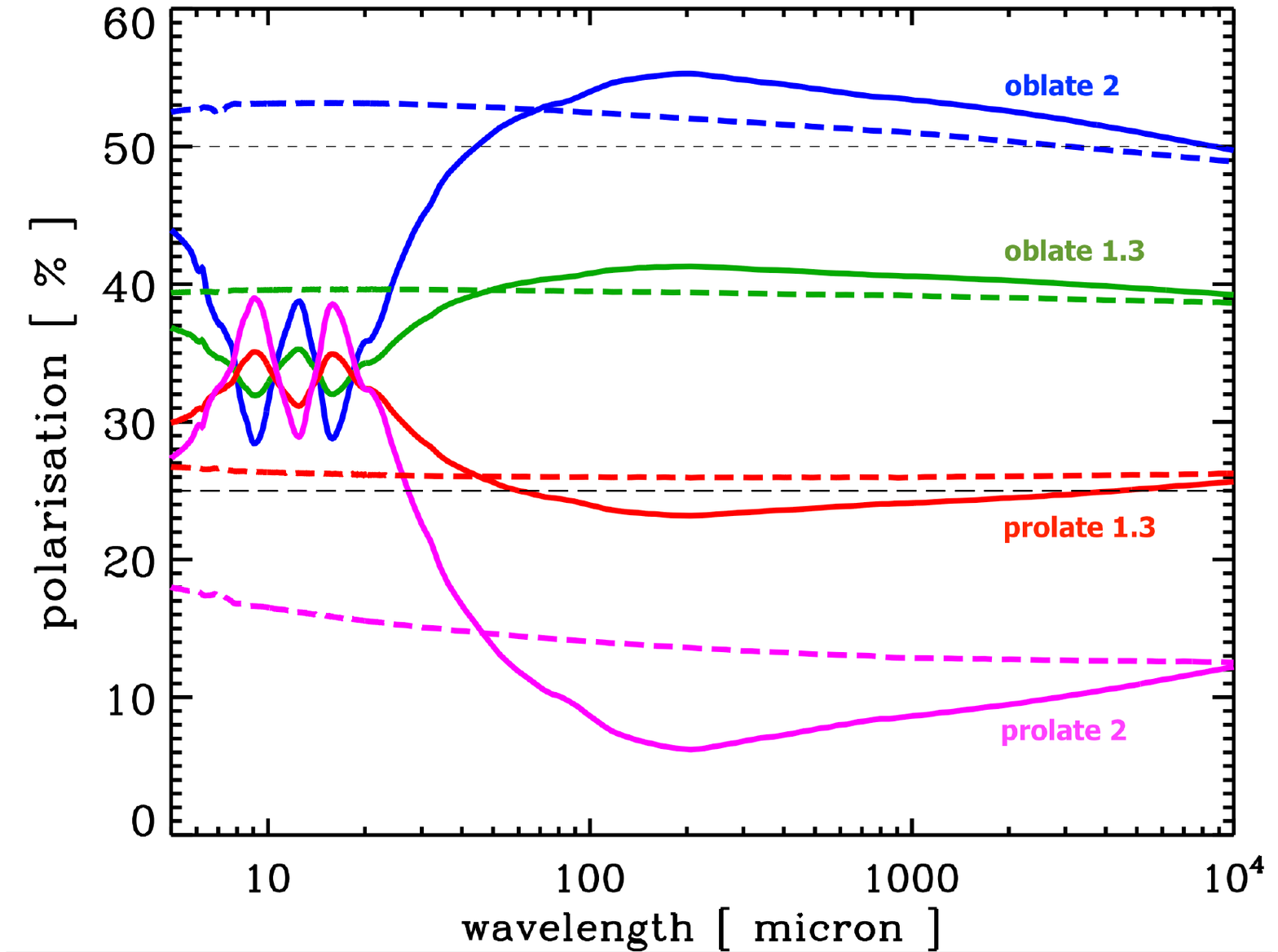}
\caption{Wavelength dependent polarisation factors, expressed as a percentage and as defined by Eq. (\ref{eq_polar_defn}), for the particles shown in Fig. \ref{fig_aSil_150} for a-Sil/a-C particles (solid lines) and in Fig. \ref{fig_aCH_150} for a-C:H/a-C particles (dashed lines) for $a_{\rm eff} =150$ and oblate particle axis ratios of 1.3 (green) and 2 (blue) and prolate particle axis ratios of 1.3 (red) and 2 (violet) for confocal mantles. The long and short dashed grey lines indicate polarisation levels of 25\% and 50\%.} 
\label{fig_polar_aSil}
\centering
\end{figure}

\subsection{Polarisation by mantled spheroids}
\label{sect_polar}

We now consider the polarisation properties of mantled spheroids. Given that the polarisation of  a particle, in extinction and absorption, is usually considered as proportional to $(\sigma_{\rm ext,\perp} - \sigma_{\rm ext,\|})/2$ and $(\sigma_{\rm abs,\perp} - \sigma_{\rm abs,\|})/2$, respectively \cite[e.g.][]{Guillet:2018hg}, we define a fractional polarisation, $f_{\rm pol}$, in extinction as 
\[
f_{\rm pol} = \frac{ \frac{2}{3} \sigma_{\rm ext,\perp} - \frac{1}{3} \sigma_{\rm ext,\|} }{ \frac{2}{3} \sigma_{\rm ext,\perp} + \frac{1}{3} \sigma_{\rm ext,\|} }
= \frac{ 2 \pi a b Q_{\rm ext,\perp} - \pi b^2 Q_{\rm ext,\|} }{ 2 \pi a b Q_{\rm ext,\perp} + \pi b^2 Q_{\rm ext,\|} }
\]
\begin{equation}
\ \ \ \ \ \ = \frac{ 2 a Q_{\rm ext,\perp} - b Q_{\rm ext,\|} }{ 2 a Q_{\rm ext,\perp} + b Q_{\rm ext,\|} }. 
\label{eq_polar_defn}
\end{equation}

\section{Results and comparisons}
\label{sect_results} 

One of the major motivations for this work was to provide an efficient means to calculate the long wavelength ($\lambda \gtrsim 8\mu$m) optical properties (extinction, absorption, scattering, and polarisation) of spheroidal CM grains within the THEMIS interstellar dust model framework\footnote{The Heterogeneous dust Evolution Model for Interstellar Solids: \  https://www.ias.u-psud.fr/themis} \citep{2013A&A...558A..62J,2017A&A...602A..46J,2014A&A...565L...9K} and its updated version THEMIS 2.0 \citep{2024A&A...684A..34Y}. We use the 10\,K amorphous silicate, a-Sil, optical constants derived by \cite{2022A&A...666A.192D} for a mix of 80\% forsterite-type and 10\% of each of the enstatite-type silicates. This is the amorphous silicate mix a-Sil-1 of THEMIS 2.0, which was adopted for the method comparisons and is not meant as a test of THEMIS 2.0.  For hydrogenated amorphous carbon, a-C(:H), we use the optical constants of \cite{2012A&A...540A...1J,2012A&A...540A...2J,2012A&A...542A..98J} and for the a-C mantle material we adopt a size-dependent effective band gap of $0.1$eV equivalent to radii of 2.5nm and 10nm, that is half of the mantle depth of 5nm on a-Sil and 20nm on a-C:H, as proposed by \cite{2013A&A...558A..62J}.\footnote{N.B. The adopted best-fit THEMIS 2.0 model actually assumes the a-Sil-2 silicate mix \citep{2024A&A...684A..34Y} and an a-C mantle material equivalent radius of 2.5nm \cite{2012A&A...542A..98J}.} Thus, and as per the THEMIS nomenclature, the considered spheroidal CM particle structures, with confocal mantles, are a-Sil/a-C and a-C:H/a-C with effective radii and mantle depths $a_{\rm eff}/d$ of 145nm/5nm and 130nm/20nm, respectively. 

The discrete dipole approximation \citep[DDA;][]{1973ApJ...186..705P} was used to calculate the grain absorption and scattering efficiencies, with version 7.3.3 of the DDSCAT routine \citep{DDA1, DDA2, DDA3}. In the DDA the grain is assumed to be accurately represented by an assembly of point-like electric dipole oscillators with the dipole size, $\delta$, set according to the following criterion: $|m|2\pi\delta/\lambda < 1/2$, where $m$ is the material complex refractive index. The absorption and scattering efficiencies must be averaged over the grain's spinning dynamics: the DDSCAT routine calculates these efficiencies by varying the grain's orientation relative to the incident electromagnetic wave. Taking advantage of the symmetries of spheroidal grains, this is simply done using 10 rotation steps each of angle of 10$^\circ$. Full details of the grain orientations and rotations are given in Sect. 3 of \citet{2024A&A...684A..34Y}.

In the following we compare the different methods (analytical, DDSCAT, and effective medium theory) by assuming a core+mantle effective radius $a_{\rm eff} = (a \, b^2)^{1/3} = [(a_c+d)(b_c+d)^2]^{1/3}$ with $d$ the mantle depth. The core and mantle volumes, $V_c$ and $V_M$ respectively, are  
\begin{equation}
V_c = (4/3) \, \pi \, (a_{\rm eff}-d)^3$ and $V_m = (4/3) \, \pi \, [ \, a_{\rm eff}^3 - (a_{\rm eff}-d)^3 \, ]. 
\end{equation}
For a given effective radius particle the core and mantle volumes are the same for each particular spheroid ellipticity and mantle depth case, that is co-axial, confocal or constant depth mantles, and also for the mantled spheres, and are therefore independent of the particle axis ratio $(a/b)$ with the effective radii normalisations. Then, given that
\[
V = \frac{4}{3} \, \pi \, a_{\rm eff} ^3 = \frac{4}{3} \, \pi \, (a \, b^2), \ \ \ \ \ {\rm for\ a\ given} \ \ \ \ \ e = (a/b) \ \ \ \ \ 
\]
\begin{equation}
{\rm it\ follows\ that} \ \ \ \ \ a = e^\frac{2}{3} \, a_{\rm eff} \ \ \ \ \ {\rm and} \ \ \ \ \ b = e^{-\frac{1}{3}} \, a_{\rm eff}.  
\end{equation}
Substituting the above in Eq.(\ref{eq_sigma_t}) the equally weighted cross-sections can generically be re-written as
\begin{equation}
\langle \sigma_i \rangle = \frac{1}{3} \, \pi \, a^2 +  \frac{2}{3} \, \pi \, ( a \, b) = \frac{\pi \, a^2}{3} \frac{1}{e} \, (e + 2) = \frac{\pi \, a_{\rm eff}^2}{3} \, e^{\frac{1}{3}} (e + 2), 
\end{equation}
where $i =$ abs or sca. Normalising to unit effective radius, that is $a_{\rm eff} = 1$, and as per Fig. \ref{fig_test_two} setting $e = \frac{1}{2}, \frac{2}{3}, 1, \frac{3}{2}$ and 2 yields $\langle \sigma_i \rangle / \pi = $ 0.66, 0.78, 1, 1.34, and 1.68, respectively. In other words, the net geometric cross-sections for equal volume and equal axis-weighted oblate and prolate spheroids are decreased and enhanced, respectively, in comparison with a mantled spherical particle with the same core and mantle volumes. Added to this, however, are the effects of the optical properties, $Q_{\rm abs}$ and $Q_{\rm sca}$, which act differentially on the total cross-sections and can reduce or amplify the magnitude of these differences depending on the parallel or perpendicular directions (see Fig. \ref{fig_test_two}). 

With thin mantles ($d < (a_c + 2 b_c)/3$), as is typically the case for CM interstellar grains, there is little discernible difference between the results for co-axial, confocal and constant depth mantles when the parallel and perpendicular factors are combined.That this must be so can be surmised from the $L_i$ factors plotted in Fig. \ref{fig_Ls} (for $a/d = 3$) because for even thinner mantles (i.e. $a/d > 30$) the differences between the core and core+mantle $L_i$ factors for all mantle forms will be considerably smaller. 

Setting the core and mantle to the same material and comparing with the exact calculations using the BHCOAT routine \citep{1998asls.book.....B} for spherical CM particles ($a = b$), yields exactly the same results. For the same materials and the same axis ratio ($a/b$) the spheroid CM calculations also give the same results as for homogeneous spheroids with $a \ne b$.

\subsection{Comparison with DDSCAT results}
\label{sect_DDA_comparison} 

In order to compare our analytical results with the numerical results of the DDSCAT code  \citep{2000ascl.soft08001D}, we do not use the $\frac{2}{3}_{\perp}$ and $\frac{1}{3}_{||}$ simplification in the cross-section calculations, as detailed above, but calculate them by integration over rotation about the principal axes. The cross-sections are fixed for spheroid rotation about the $a$ and $c^\prime$ axes, that is $\sigma_i = \pi a b$ in both cases, and therefore only integration about the $b$ axis is required. The cross-section integration is ideally performed as follows 
\[
\langle \sigma_i \rangle = \frac{1}{3} \left[ \ 2 \ \sigma_{\perp,i}(ab) \, Q_{\perp,i}(ab) \ +  \int_0^{\pi/2} \sigma_{||,i}(\theta) \, Q_{i}(\theta) \ d\theta \ \right]  \ \ \ \ \ {\rm and} \ \ \ \ \ 
\]
\begin{equation}
\sigma_{||,i}(\theta) = \pi \ b \ \surd [ a^2 {\rm cos}^2 \theta + b^2 {\rm sin}^2 \theta ] = \pi \ b \,A,
\end{equation}
where $A \ = \surd [ a^2 {\rm cos}^2 \theta + b^2 {\rm sin}^2 \theta ]$ is the projection of the semi-axis $a$ onto the $ac^\prime$ plane (that is for $\theta = 0$). The term $\sigma_{||,i}(\theta)$ takes the expected limiting values of $\pi a b$ for $\theta = 0$ and $\pi b^2$ for $\theta = \pi / 2$. This exact form of the integral requires that the efficiency factor $Q_{i}(\theta)$ is known for any arbitrary angle $\theta$. However, for the analytical method adopted here $Q_{i}$ is only determinable in the orthogonal $\perp$ and $||$ directions (see above), and so the cross-section at each angle step is taken to be an angle-weighted sum of the $Q_{\perp,i}$ and $Q_{||,i}$ factors. This is achieved by replacing the integral with a sum over $N_\theta$ steps each of weight $N_\theta^{-1}$ as follows
\[
\sigma_{||,i}(\theta) = \frac{1}{N_\theta} \sum_0^{\pi/2} \left[ \, f_a  \, Q_{\perp,i} + f_b \, Q_{||,i} \, \right]   \ \ \ \ \ 
\]
\begin{equation}
{\rm where} \ \ \ \ \ f_a =  \left( \frac{A}{b} - 1 \right)\ \ \ \ \ {\rm and} \ \ \ \ \ f_b = ( 1 - f_a ). 
\end{equation}
We find that 40 rotation angle steps about the $b$ axis ($N_\theta = 40$) are sufficient to achieve satisfactory convergence. In the DDSCAT calculations only 10 angle steps about the $b$ axis were taken due to computational time limitations, which accounts for some small differences in the comparison results but otherwise has little effect. 

We compared the mantled spheroid analytical method results with those generated by the DDSCAT tool  \citep{2000ascl.soft08001D} for particles with sufficient volume resolution to ensure a close representation of the exact spheroidal forms.  For coated spheres we can also compare with the Mie routine BHCOAT provided by \cite{1998asls.book.....B}. In the case of spherical CM particles the three methods agree to within 3\% over the $5-300\mu$m wavelength range and to within 10\% out to centimetre wavelengths. For spheroidal CM particles in a comparison of the analytical results with those of DDSCAT, as presented in Figs. \ref{fig_aSil_150} an \ref{fig_aCH_150}, we found that the shortest wavelength to which the long wavelength analytical method can reliably be applied is $\lambda_{\rm lim} \simeq 8\mu$m and good for all $\lambda > 8\mu$m (see Figs. \ref{fig_aSil_150} and \ref{fig_aCH_150}). 

Given that the THEMIS large silicate and carbon grain log-normal distributions peak at radii close to 140 and 160nm, respectively \citep{2017A&A...602A..46J}, we use 150nm as a standard test radius and consider 5nm and 20nm confocal a-C mantles on a-Sil and a-C:H cores, as used in our DDSCAT calculations. The DDSCAT results and those for the long wavelength analytical method, for $\lambda = 1-1000
\mu$m, for 150nm CM grains of a-Sil/a-C are shown in Fig. \ref{fig_aSil_150} and those for a-C:H/a-C shown in Fig. \ref{fig_aCH_150}. We note that while the analytical method provides exact solutions for core-mantle spheroids DDSCAT can, at best, only ever provide good approximations to these perfect shapes. We find intrinsic discrepancies between DDSCAT and the exact analytical method of the order of 10\% overall; typically 3 to 6\% for the a-Sil/a-C and mostly within 4 to 15\% (always $\leqslant 26$\%) for the a-C:H/a-C CM particles.\footnote{Strictly, the  differences  between the results of the two methods cannot be considered as errors. On the one hand, the method presented here assumes perfect forms to model what are, in reality, the less than perfect shapes of interstellar grains and atmospheric aerosols. On the other hand the digitised and granular forms of the DDSCAT particles that approximate perfect spheroids are perhaps more realistic.} In most cases the differences are relatively small but do increase at millimetre to centimetre wavelengths. The differences between the methods for other thin-mantled a-Sil/a-C grains are similar in magnitude and do not depend upon axis ratio ($a/b$), or oblate or prolate form. For the a-C:H/a-C grains the differences are about a factor of 2 larger. 

The agreement between the analytical and DDSCAT results is better for thin-mantled silicate spheroids than it is for thick-mantled carbon grains. The discrepancies for the latter spheroids increase with increasing wavelength, which may arise because the thicker mantles are more coarsely discretised in the DDSCAT calculations and do not sufficiently resolve the mantles. For example, mantle depth to radius ratios of 0.07--0.4  can impose prohibitively time-consuming computations.\footnote{For example, for DDSCAT to sufficiently resolve 5nm (20nm) deep mantles on particles with $a_{\rm eff} = 150$nm generally requires 3 (12) volume elements (dipoles) across the mantle and therefore 87 (78) across the core radius, because the dipole size is fixed throughout, requiring a total of $\approx 3$ million dipoles. If 3 volume elements are used to resolve the 20nm mantle then only $\approx 5 \times 10^4$ dipoles are required in all but the approximation of a smooth CM structure is then poor.} 

We note that the addition of fixed depth a-C mantles onto a-Sil cores essentially adds a constant slope in a log-log plot, with a power law index of $\simeq 1-1.5$, to the silicate $Q_{\rm abs}$, an effect that is obviously more pronounced for smaller particles. The net result is that of a diminution of silicate features in the a-Sil/a-C spheroids through a reduced band-to-continuum contrast, which will be more pronounced the smaller the spheroid because of the increased mantle volume with respect to the core.

Fig. \ref{fig_polar_aSil} show the polarisation behaviour of the same particles shown in Figs. \ref{fig_aSil_150} and \ref{fig_aCH_150}.  From Fig. \ref{fig_polar_aSil} it can be seen that, for a-Sil/a-C particles with 5\,nm a-C mantles the polarisation is rather flat from FIR to centimetre wavelengths, while for a-C:H/a-C particles with 20\,nm mantles the quasi-flat polarisation behaviour extends down to IR wavelengths. 
The degree of polarisation over these same wavelength ranges increases with the transition from prolate to oblate form 
with polarisations $\simeq 10, 25, 40$, and 50\% for $a/b = 2, 1.3, 0.8$, and 0.5, for prolate 2, prolate 1.3, oblate 1.3, and oblate 2, respectively, with only relatively small differences between the two very different particle compositions. 

It therefore appears that the degree of polarisation for both a-Sil and a-C:H a-C mantled grains is more sensitive to the particle shape than it is to the core composition or the mantle thickness. 

\subsection{Comparison with volume-normalised coated spheres}
\label{sect_CS_comparison} 

Within the framework of the THEMIS model it is useful and interesting to have some idea of the likely effects of non-sphericity on CM grain particles. In order to estimate this we compare the exact Mie calculations for mantled spheres and spheroids with the same core and mantle volumes. The black dashed lines in Figs. \ref{fig_aSil_150} and \ref{fig_aCH_150} show the results for the mantled spheres. For the a-C carbon-mantled a-Sil and a-C:H particles the coated sphere results closely follow those of the mantled spheroids with no discernible differences in the wavelength-dependent behaviour. Thus, and apart from furnishing no polarisation data, mantled spheres would appear to provide a good representation of the optical properties of carbon-mantled spheroids.

\subsection{Comparison with a volume-normalised, effective medium, volume averaging method}
\label{sect_EMT_comparison} 

A commonly used technique in deriving the optical properties of bi-component dust particles, or even multi-coated particles \citep{2011A&A...528A..98J},  is to use an effective medium theory to average the complex dielectric function of the components. In this context, perhaps the most useful is the Bruggemann effective medium theory \citep[EMT,][]{Bruggemann1935}, which derives an effective dielectric function, $\epsilon_{av}$, for a fractional volume of inclusions $f_i$ of dielectric function $\epsilon_i$ in a matrix of of dielectric function $\epsilon_m$. The effective dielectric function is found by solution of the Bruggemann equation  
\begin{equation}
f_i \left[ \frac{\epsilon_i - \epsilon_{av}}{\epsilon_i + 2 \epsilon_{av}} \right] + (1-f_i) \left[ \frac{\epsilon_m - \epsilon_{av}}{\epsilon_m + 2 \epsilon_{av}} \right] = 0.
\label{eq_BEMT}
\end{equation}
Unlike the Maxwell-Garnett EMT \citep{1904RSPTA.203..385G} the Bruggemann equation is symmetric about the interchange of inclusions and matrix. For a binary mix Eq.(\ref{eq_BEMT}) requires solving a quadratic equation or higher order polynomials for more than two components \cite[e.g.,][]{2011A&A...528A..98J}. However, the Bruggemann equation can be applied to more than binary mixtures, as pointed out by \cite{2011A&A...528A..98J}, if the equation is applied recursively starting with the most refractory of the component materials and proceeding to the least refractory in a pair-wise manner. 

The results for Bruggemann EMT-averaged homogeneous spheroids are shown by the thin and thick green lines in Figs. \ref{fig_aSil_150} and \ref{fig_aCH_150}. The EMT, DDSCAT, and analytical method results compare well for a-Sil/a-C particles (Fig. \ref{fig_aSil_150}) over a wide range of radii ($a_{\rm eff} = 50-300$, only the 150nm data are shown here). Thus, we find a good agreement between all of the methods for thin absorbing mantles on spheroidal refractory amorphous silicate cores, where the mantle/(core+mantle) volume ratio is 0.27, 0.10, and 0.05, for $a_{\rm eff} = 50, 150$, and 300nm, respectively. For the 150nm a-C:H/a-C particles (Fig. \ref{fig_aCH_150}) we find that the absorption efficiency factor, $Q_{\rm abs}$, is underestimated by at least 50\% for the EMT method. For the equivalent 50nm particles the $Q_{\rm abs}$ (not shown) is generally good to within $\pm 10$\%. However, for 300nm radius grains $Q_{\rm abs}$ is systematically underestimated by an order of magnitude or more at millimetre to centimetre wavelengths. For these particles the mantle/(core+mantle) volume ratio is  0.78, 0.35 and 0.19, for $a_{\rm eff} = 50, 150$, and 300nm, respectively. Thus, it would seem that in the spheroidal a-C:H/a-C CM EMT calculations, for a fixed depth (20nm) absorbing mantle, the discrepancy increases significantly with increasing radius, with the absorption properties always being systematically underestimated by the EMT method. 

An explanation for the unsatisfactory EMT results for thick absorbing mantles on grains, which must also equally well apply to spherical as well as spheroidal particles, lies in a dilution effect that leads to an underestimate of the effects of the mantle contribution to the optical properties in absorption. The averaging of the dielectric functions of the component parts of a particle to yield an effective dielectric function was designed to apply to intimate but distinct component mixes, such as inclusions within a matrix, rather than more macroscopically distinct CM structures. When applied to the latter structures the dielectric function of the mantle is diluted into that of the core which may have a very different dielectric function.\footnote{Such as semi-conduting mantles on insulating cores, as per the large a-Sil/a-C and a-C:H/a-C THEMIS CM grains.} What this does is to underestimate the effects of a surrounding mantle, which in a CM particle may dominate the optical properties and effectively shield the core. This effect explains why, for a fixed mantle depth, the underestimate of the absorption properties increase with size. This is due to an increasing dilution of the core dielectric function into that of the core, with increasing size, in any effective medium treatment. 

The Bruggemann EMT and Maxwell-Garnett EMT methods should therefore not be applied to the case of thick, absorbing mantles on refractory cores. More generally, EMT methods will always give unreliable results in the case of CM particle structures, be they spheres, spheroids or of any arbitrary shape, where there is a large difference in the dielectric functions of the core and the mantle materials.

\section{Summary and concluding remarks}
\label{sect_conclusions}

The analytical methods developed here to derive the long wavelength optical and polarisation properties of coated spheroidal particles, of arbitrary semi-axis ratio, provide exact mathematical solutions for their extinction, absorption, scattering, and polarisation properties. These same properties can be calculated with, albeit more time consuming, numerical methods such as DDSCAT \citep{2000ascl.soft08001D}, which have been used to approximate these same mathematical shapes using volume element discretisation. However, interstellar grains are quite clearly not exact or perfect mathematical figures and so comparison with the DDSCAT results can be used to give us some idea about the likely uncertainties that should be associated with the analytically-calculated spheroid optical properties as applied to real interstellar grains. One obvious advantage of the heavier numerical methods is that, unlike the method developed here, they can be used to study arbitrary shapes, structures and  compositions down to UV wavelengths. 

In the analysis presented here we find good agreement between the analytical and DDSCAT methods, for wavelengths $\geqslant 8\mu$m,  for 5nm carbon-mantled (a-C) amorphous silicate particles (a-Sil), with differences generally less than $\pm 6$\% from IR to centimetre wavelengths, while for hydrogenated amorphous carbon particles with 20nm aromatic-rich a-C mantles on aliphatic-rich a-C:H cores the differences are about a factor of 2 larger. In both cases, the differences tend to increase with increasing wavelength. 
Thus, the analytical method developed here and the numerical DDSCAT calculations show good agreement for the particle absorption and scattering properties of CM particles. The discrepancies between the exact analytical method for perfect spheroids and that for the more irregular voxelated spheroids imply that the uncertainties in the optical properties used by dust models are of the order of $\pm 10$\% overall; typically within a few percent at IR wavelengths but up to $\pm 25$\% at millimetre to centimetre wavelengths. These are the intrinsic uncertainties due to the inexactitudes in modelling the unknown interstellar grain shapes and structures within the framework of any particular dust model. In this sense the discrepancies between the studied methods translate into the uncertainties inherent in modelling the irregular nature of real interstellar grains and it is unlikely that we can realistically do any better than this. 

The new analytical method for spheroidal CM particles has the major advantage over numerical methods, such as DDSCAT and T-Matrix, in that it is easy to code and is computationally fast. It can therefore be used to quickly explore a wide parameter space encompassing grain composition, size, spheroidal shape, and mantle composition, form, and depth at IR to centimetre wavelengths ($\lambda = 8 - 10^4 \mu$m). The resulting optical property and polarisation predictions will be useful in investigating and modelling the dust properties in the diverse diffuse interstellar and circumstellar media in the Milky Way and in extragalactic environments, but may be particularly pertinent in the interpretation of the dust polarisation in planet forming circumstellar disk observations where CM-type particle structures might be expected. 

The analytical method presented here compares extremely well with the coated sphere results derived using the Mie theory, implying that CM spherical particles can be used as a good approximation for CM spheroidal particles. However, this new analytical method has the advantage that it can also be used to calculate the polarisation properties of spheroidal particles. 

We have compared the analytical and DDSCAT methods with the Bruggemann EMT dielectric function averaging method and find that for thin mantled particles there is reasonable agreement. However, for thick absorbing mantles the EMT results underestimate the absorption properties by large factors and should therefore not be used in this case. The cause of the large errors is the dilution of the critical absorbing mantle properties into an effective or averaged dielectric function. 

We caution against the use of EMT methods to approximate the optical properties of CM particles, of any shape, where the mantles make up a significant fraction of the total volume and where there is a large difference in the core and mantle dielectric functions. For absorbing mantles the absorption properties will always be underestimated and for absorbing cores they will always be overestimated. Thus, EMT methods are best used with a great deal of caution in the study of particles consisting of cores with encompassing mantles. 

\begin{acknowledgements}
This work would not have been possible were it not for the shoulders of giants. 
\end{acknowledgements}

\bibliographystyle{aa} 
\bibliography{Ant_bibliography}


%
%

\end{document}